\documentclass[12pt]{iopart}
\usepackage{iopams}  
\usepackage{graphicx}

\begin{document}

\title[]{Some remarks on `superradiant' phase transitions in light-matter systems}

\author{Jonas Larson}
\address{Department of Physics, Stockholm University, Se-106 91
  Stockholm, Sweden}
\author{Elinor K. Irish}
\address{Physics and Astronomy, University of Southampton, High eld, Southampton, SO17 1BJ, UK}
  
\begin{abstract}
In this paper we analyze properties of the phase transition that appears in a set of quantum optical models; Dicke, Tavis-Cummings, quantum Rabi, and finally the Jaynes-Cummings model. As the light-matter coupling is increased into the deep strong coupling regime, the ground state turns from vacuum to become a superradiant state characterized by both atomic and photonic excitations. It is pointed out that all four transitions are of the mean-field type, that quantum fluctuations are negligible, and hence these fluctuations cannot be responsible for the corresponding vacuum instability. In this respect, these are not quantum phase transitions. In the case of the Tavis-Cummings and Jaynes-Cummings models, the continuous symmetry of these models implies that quantum fluctuations are not only negligible, but strictly zero. However, all models possess a non-analyticity in the ground state in agreement with a continuous quantum phase transition. As such, it is a matter of taste whether the transitions should be termed quantum or not. In addition, we also consider the modifications of the transitions when photon losses are present. For the Dicke and Rabi models these non-equilibrium steady states remain critical, while the criticality for the open Tavis-Cummings and Jaynes-Cummings models is completely lost, {\it i.e.} in realistic settings one cannot expect a true critical behaviour for the two last models.
\end{abstract}

% Uncomment for PACS numbers
%
% Uncomment for keywords
%\vspace{2pc}
%\noindent{\it Keywords}: XXXXXX, YYYYYYYY, ZZZZZZZZZ
%
% Uncomment for Submitted to journal title message
%\submitto{\JPA}
%
% Uncomment if a separate title page is required
\maketitle
% 
% For two-column output uncomment the next line and choose [10pt] rather than [12pt] in the \documentclass declaration
%\ioptwocol
%

\section{Introduction}
The {\it deep strong coupling} regime~\cite{dsc} of light-matter models is characterized by an instability of the photon vacuum at zero temperature. The simplest realization is found in the quantum Rabi model~\cite{rabi} describing a two-level `atom' dipole interacting with a single photon mode. Also the many-atom version of the Rabi model, the Dicke model~\cite{dicke}, displays the same instability. The companion models derived by applying the {\it rotating wave approximation} (RWA)~\cite{rwa} to the Rabi and Dicke models, the Jaynes-Cummings~\cite{jcm} and Tavis-Cummings~\cite{tcm} models, also possess a ground state different from the vacuum in the deep strong coupling regime. Taking the proper thermodynamic limits, all four models become critical~\cite{DickePT,DickePT2,TCPT,rabipt1,rabipt2,plenio1,plenio4}, {\it i.e.} the crossover turns into a true phase transition (PT) for a certain critical coupling $g_c$. That is, for $g<g_c$ the field is in the vacuum plus the atoms in their lower electronic state and for $g>g_c$ both the field and the atoms become excited. The low coupling phase is the symmetric {\it normal} one, while the strong coupling phase is the symmetry broken {\it superradiant} one. This kind of {\it normal-superradiant} (NS) PT was first discovered in the TC model~\cite{TCPT}, but soon afterwards it was shown that it survives also in the Dicke model where the RWA has not been applied~\cite{DickePT,DickePT2}. Here, the thermodynamic limit is identified as letting the particle number go to infinity which can be seen as the classical limit for the atomic (spin) part. Recently it was discovered that a PT also appears for only a single atom (spin-1/2) provided that the classical limit of the oscillator is taken instead~\cite{rabipt1,rabipt2,plenio1,plenio4}. 

These models are all quantum, and since the transition occurs at zero temperature~\cite{zero}, normally the transitions have been considered as {\it quantum phase transitions} (QPTs). For a QPT, the instability is not driven by thermal fluctuations (we assume $T=0$), but rather by quantum fluctuations~\cite{sachdev}. A QPT necessarily implies that the ground state energy $E_0(g)$ as a function of the coupling $g$ shows a non-analytic behaviour at the critical point. In particular, for a {\it first order} QPT the derivative $\partial_g E_0(g)$ is discontinuous at $g_c$, while for a continuous QPT $\partial_g^2 E_0(g)$ is discontinuous. Taking these non-analyticity of $E_0(g)$ as a defining property of a QPT, the four models support a QPT in the thermodynamic limit. However, you may have a non-analyticity in the ground state without having an instability driven by quantum fluctuation. The Hamiltonian $\hat H=g\hat\sigma_z$ with $\hat\sigma_z$ the $z$-Pauli matrix is the simplest example of such a situation; the eigenenergies cross at $g=0$. For quantum fluctuations to become influential we need that different constituents of the Hamiltonian do not commute among themselves. 

In the present paper we analyze this aspect in terms of the models mentioned above. And we show that neither of the transitions are driven by quantum fluctuations. For the Rabi and Dicke models the fluctuations are vanishingly small in the thermodynamic limit, and for the TC and JC models the fluctuations are strictly zero at any system size. Indeed, all transitions are of the mean-field type meaning that the mean-field predictions are exact. 

Any other fluctuations may trigger the transition and especially those stemming from a coupling to an environment. This also motivates studying the corresponding PT for the open models. A further reason to do so is that experimental realizations are inevitably open, and also driven in order to reach the deep strong coupling regime. To give a strict definition of criticality in such open-driven systems one considers the system's steady state, and explores whether it is non-analytic. It is known that the open Dicke model (and also the open Rabi model) is critical, but here we show, both numerically and analytically, that in contrast the open TC and JC models are not critical. The unique steady state of these two models, regardless of parameters, is the vacuum plus all atoms in their lower states. 

The paper is organized as follows. In the next section we give a very brief introduction to PT's and QPT's. We mention the phenomena needed for later sections, like universality and critical exponents. We also give a well known example that shares some properties with the superradiant PT's of the Dicke and Rabi models discussed in the following section. Also the idea of non-equilibrium PT's is summarized in Sec.~\ref{sec2}. Section~\ref{sec3} contains the results on the four models in different subsections. Finally in the last section we summarize the results and provide some further discussions. In addition, in the appendix we present another paradigmatic example known from condensed matter theory. This example serves as a comparison to the TC and JC models, there are evident similarities but also qualitative differences. 

\section{Quantum phase transitions -- a prelude}\label{sec2}
\subsection{Equilibrium transitions}
Before discussing the NS transition in general terms we need to recapitulate some basic theory of critical phenomena. By now, the physics of equilibrium PT's is well developed and understood, especially with the insight provided by the {\it renormalization group techniques}~\cite{RG}. The key point here is that of {\it scale invariance}, meaning that for a {\it continuous} PT (also called {\it second order} PT), at the critical point the characteristic length scale diverges as
\begin{equation}\label{critexp1}
\xi\propto|g-g_c|^{-\nu}.
\end{equation}
Here, the model Hamiltonian $\hat H(g)$ depends on some coupling parameter $g$ and $g_c$ is the critical coupling at which the system becomes critical, and $\nu$ is the {\it correlation length critical exponent}. The critical point distinguishes two phases differing in their physical properties, and the point marks a non-analyticity in the system state and thereby also in various measurable quantities. The behaviour (\ref{critexp1}) implies that the physics in the vicinity of the critical point cannot depend on microscopic details, like the lattice spacing, but it relies solely on macroscopic properties, {\it i.e.} symmetries and dimensionality. This is the concept of {\it universality}; the critical exponent $\nu$ determines how the correlations behave close to the critical point $g_c$ (the proportionality constant is not universal and may depend on microscopic details). There is a set of critical exponents and their values define the {\it universality class} that the system belong to. In addition, there exist also a set of {\it scaling laws} (based on some general arguments and observations) that constrain the critical exponents such that they are not all independent from one another. For a classical phase transition we may have that $g$ represents the temperature $T$ such that upon lowering $T$ there exists a critical temperature $T_c$ for which the system goes from a disordered phase into an ordered one characterized by {\it long range order}. For the system to support long range order we must have that the thermal fluctuations cannot be too strong, which implies that PT's are not always allowed. This is the result of the {\it Mermin-Wagner theorem}~\cite{AA} which states that in lower dimensions (one and two) transitions are forbidden or a discrete symmetry has to be broken. The theorem can be applied to QPT's as well, but the dimensionality is then changed according to the {\it quantum--classical mapping}~\cite{sachdev}. Another feature of scale invariance is that of a diverging time-scale at the critical point,
\begin{equation}\label{critexp2}
\tau\propto|g-g_c|^{-\nu z},
\end{equation}
where we have introduced the {\it dynamical critical exponent} $z$. Since the inverse time defines an energy we also have a vanishing characteristic energy $\Delta\propto|g-g_c|^{\nu z}$ at the critical point. As we will argue below, for a QPT $\Delta$ is the energy gap to the first excited state.

\begin{figure}[h]
\centerline{\includegraphics[width=8cm]{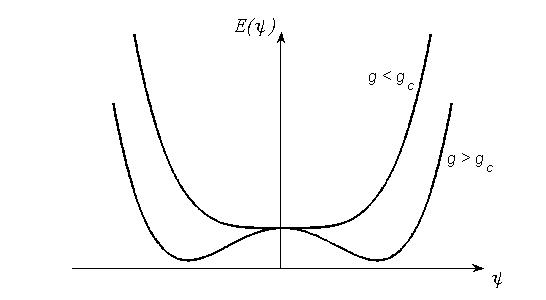}}
\caption{The idea of Landau mean-field theory and symmetry breaking. Below the critical coupling the energy functional has a global minimum at $\psi=0$, and above it builds up two minima symmetrically around the origin. Symmetry breaking implies that the system, of random, picks one of these solutions. In a classical PT, the solution at $\psi=0$ for $g>g_c$ is destabilized due to thermal fluctuations, while for a QPT the fluctuations causing the destabilization derive from the Heisenberg uncertainty.} \label{fig1}
\end{figure}

The non-analyticity is, as pointed out, manifested in observables and in particular we characterize a given phase by an {\it order parameter} $\psi$. The order parameter carries physical information about the state and obeys
\begin{equation}\label{order}
\psi=\left\{
\begin{array}{lll}
0, & & g<g_c,\\ 
\neq0, & & g>g_c.
\end{array}\right.
\end{equation} 
Universality implies that in the so called {\it symmetry broken phase} $\psi\propto|g-g_c|^\beta$ for a new critical exponent $\beta$. Within a {\it mean-field approximation} we find the value of the order parameter by optimizing some energy functional $\mathcal{E}[\psi]$. In {\it Landau theory} it is the Landau free energy that is omptimized~\cite{RG}, and we will see in the following sections when discussing the NS transition that one of our mean-field treatments consists in optimizing an effective semi-classical Hamiltonian. Stability of the solution implies that the optima should be local minima. Thus, in the {\it symmetric phase} where $\psi=0$ the energy functional should possess a global minimum at $\psi=0$, while in the other phase, the symmetry broken phase, the minimum should be for $\psi\neq0$. As $\psi$ represents an actual observable we have that $\mathcal{E}[\psi]$ should obey the same type of symmetries as the actual Hamiltonian. This means that if $\psi$ is a solution, so will $-\psi$ be provided that the model supports a parity symmetry, {\it i.e.} a $\mathbb{Z}_2$ symmetry. This situation is schematically presented in Fig.~\ref{fig1}. Above the critical coupling $g_c$ the functional attains a `double-well' structure. There is still a solution $\psi=0$ also for $g>g_c$ but this is not energetically stable meaning that fluctuations (thermal or quantum for a classical PT or QPT respectively) will cause the system to pick one of the two global minima. This is the result of {\it spontaneous symmetry breaking}; your solution does not share the symmetry as your Hamiltonian (in this schematic example the $\mathbb{Z}_2$ parity symmetry). If the system supports instead a continuous symmetry one has that $\mathcal{E}[\psi]=\mathcal{E}[\exp(i\phi)\psi]$ for any angle $\phi$ and $\mathcal{E}[\psi]$ now lives in the complex $\psi$-plane, and in particular, for $g>g_c$ it has the typical sombrero shape. 

At the quantum level, when quantum fluctuations drive the transition, one may argue that the ground state must carry the same parity symmetry as the Hamiltonian, hence it should be the symmetric superposition. The only possibility for a symmetry breaking to occur is when the ground state becomes degenerate, {\it e.g.} for the $\mathbb{Z}_2$ symmetric case the symmetric ground state and the anti-symmetric first excited state become degenerate. Such a degeneracy implies that the energy barrier separating the two minima in Fig.~\ref{fig1} prevents any tunneling between the wells -- it is infinitely high/wide. This can only happen in the {\it thermodynamic limit} when the system becomes infinite; the number of particles goes to infinity at the same rate as the system volume grows. Another result is that in the symmetry broken phase, modes of the system become macroscopically populated, typically $|\psi|\gg0$.     

Naturally, the appearance of a ground state degeneracy is directly reflected in the characteristics of the spectrum. Furthermore, the non-analyticity of the order parameter is associated with non-analytic behaviour of the ground state $|\psi_0(g)\rangle$ at $g=g_c$. The ground state energy $E_0(g)$ is a continuous function of $g$, and criticality manifests in the derivatives of $E_0(g)$. If the first derivative $\partial_gE_0(g)$ is discontinuous at $g=g_c$ we identify a {\it first order} PT, while if $\partial_g^2E_0(g)$ is discontinuous we have a continuous PT. Generally, we can decompose the system Hamiltonian as 
\begin{equation}
\hat H=\hat H_0+g\hat H_1. 
\end{equation}
The two limiting cases $g=0$ and $g\rightarrow\infty$ tell us that the two phases are dominated by $\hat H_0$ and $\hat H_1$ respectively, and especially interesting is whenever the two support different symmetries. At the critical point there is no clear cut between dominant and sub-dominant parts. We should have that for some $g$, the ground state energies of the two Hamiltonians, $\hat H_0$ and $\hat H_1$, cross. If $[\hat H_0,\hat H_1]=0$ this is a true crossing, but if $[\hat H_0,\hat H_1]\neq0$ the crossing is avoided. For the non-analyticity to emerge we must have that the gap at the crossing should close, which again is only possible in the thermodynamic limit. Thus, it is possible to have gap closing and still non-commutability $[\hat H_0,\hat H_1]\neq0$. In this case the fluctuations driving the transition are quantum and stem from the Heisenberg uncertainty. We already pointed out that the gap closing is given by the critical exponent $\nu z$, and that in the symmetry broken phase the ground state is degenerate. A further general feature is that of the excitations in this phase. If the symmetry that is broken is discrete the spectrum is gapped, while for breaking of a continuous symmetry it is gapless and you find a continuum of energies. These are the so called {\it Higgs} and {\it Goldstone modes} respectively. In a magnet, for example, the excitations in a gapped (Higgs) system are {\it kinks}/{\it domain walls} and for a gapless (Goldstone) they are {\it spin waves}~\cite{AA}. 

As a demonstration of the above discussion of QPT's we consider a paradigmatic example, that will also serve for later comparisons with the NS PT's. The Hamiltonian for the one dimensional {\it Ising model} in a transverse field is given by
\begin{equation}\label{ising}
\hat H_\mathrm{I}=-\sum_i\left(\hat\sigma_i^z+g\hat\sigma_i^x\hat\sigma_{i+1}^x\right).
\end{equation}
Here, the model is one dimensional, and the thermodynamic limit consists in letting the lattice index $i$ run over all integers. Quantum fluctuations stem from the non-commutability among the Pauli matrices $\left[\hat\sigma_i^\alpha,\hat\sigma_j^\beta\right]=i\delta_{ij}\varepsilon_{\alpha\beta\gamma}\hat\sigma_i^\gamma$ with $\delta_{ij}$ the Kronecker delta, $\varepsilon_{\alpha\beta\gamma}$ the Levi-Civita symbol, and $\alpha,\,\beta,\,\gamma=x,\,y,\,z$. In particular, the `field' $\sum_i\hat\sigma_i^z$ does not commute with the superexchange part proportional to $g$. Important for the present work is to appreciate the fact that the role played by quantum fluctuations is `size independent', in the sense that at every site the local Hilbert space is finite regardless of lattice size. Hence, quantum fluctuations cannot be disregarded in the thermodynamic limit. Furthermore, we note that the Ising model hosts a discrete $\mathbb{Z}_2$ symmetry (a $\pi$-rotation around the $\hat\sigma_i^z$'s). 

The model is most easily diagonalized by a {\it Jordan-Wigner transformation} that maps the spins into spinless fermions~\cite{sachdev}. After a {\it Bogolluobov} and Fourier transform, one diagonalizes the Hamiltonian as
\begin{equation}\label{ising2}
\hat H_\mathrm{I}=\sum_k\varepsilon_k\left(\hat c_k^\dagger\hat c_k-\frac{1}{2}\right),
\end{equation}
with the dispersion 
\begin{equation}\label{ising3}
\varepsilon_k=2\left(1+1/g^2-2\cos(k)/g\right)^{1/2}.
\end{equation}
The $\hat c_k$ and $\hat c_k^\dagger$ are fermion creation and annihilation operators and $k\in(-\pi,+\pi]$ is the {\it quasi momentum}. Since $\varepsilon_k$ is non-negative we have that the ground state is the vacuum $|0\rangle$ with no fermions at all. Thus, the ground state energy is
\begin{equation}
E_0(g)=-\frac{1}{2}\sum_k\varepsilon_k=\left(2+\frac{2}{g}\right)\mathrm{Erf}\left(\frac{4g}{(1+g)^2}\right),
\end{equation}
where we have turned the sum into an integral in the last step, and $\mathrm{Erf}(x)$ is the error function. Excitations are given by acting with the $\hat c_k^\dagger$ operators on the vacuum. The first excitation is thus given by $\hat c_0^\dagger|0\rangle$ with the excitation energy $\Delta=\varepsilon_0=2|1-1/g|$. We see that the gap closes at the critical point $g_c=1$ as $\Delta\propto|g-g_c|$ such that the critical exponent $\nu z=1$. On either side of the critical point the spectrum is gapped in accordance with the predictions for a Higgs mode. Mean-field predicts a critical exponent $\beta=1/2$ for the polarization, $\langle\hat\sigma_x\rangle\propto|g-1|^{1/2}$ of the transverse Ising model. A full quantum treatment gives, however, $\beta=1/8$~\cite{sachdev}, and thus, this is an example of a low dimensional model where the quantum correlations are significant and cannot be disregarded. Another example of a one dimensional spin chain, but supporting a continuous symmetry, the Heisenberg $X\!X$ model, is discussed in the Appendix.

\subsection{Non-equilibrium transitions}
In most light-matter quantum optical experiments, especially photon losses are inevitable. Thermodynamics teaches us that a system in contact with a thermal reservoir will eventually thermalize with the same temperature as its reservoir. So for a zero temperature reservoir the steady state of the system is its ground state. Nothing prevents, however, that the system reaches a different non-equilibrium steady state when the system itself is driven and/or the couplings to the reservoir are engineered beyond a standard dissipation exchange. With the possibility to prepare non-equilibrium steady states $\hat\rho_\mathrm{ss}$ of the system, different from thermal ones, new scenarios open up where PT's may occur.

We will be concerned with situations where the effect of the reservoir can be captured with the {\it Lindblad master equation} formalism~\cite{bruer}. In experiments operating in the optical regime, the approximations associated with such master equations are typically justified. The general form of the Lindblad master equation reads
\begin{equation}\label{lindblad}
\partial_t\hat\rho=\hat{\mathcal L}_{\hat H,\,\hat L_i}\left[\hat\rho\right]\equiv i\left[\hat\rho,\hat H\right]+\sum_i\kappa_i\left(2\hat L_i\hat\rho\hat L_i^\dagger-\hat L_i^\dagger\hat L_i\hat\rho-\hat\rho\hat L_i^\dagger\hat L_i\right).
\end{equation}  
The system Hamiltonian $\hat H$ includes in principle a {\it Lamb shift}, but this is not of importance for us and we will consider it to be the `bare' Hamiltonian of the system in the absence of the reservoir\footnote{To be precise, the Hamiltonians for our models are given in an interaction frame rotating with the frequency of the drive.}. The forms of the `jump operators' $\hat L_i$ are rather general but depend on the actual engineered system-reservoir coupling. The corresponding decay rates are $\kappa_i$. 

PT's for these driven-dissipative systems are defined as a non-analyticity in the steady state solutions $\hat\rho_\mathrm{ss}$ of Eq.~(\ref{lindblad})~\cite{openPT1,openPT2}. PT's in this respect may no longer only derive from competing terms in the Hamiltonian, but as an interplay between the drive and the dissipation. Note that in the absence of a reservoir, $\kappa_i=0$ for all $i$, the ground state of the Hamiltonian is a steady state and the present definition for PT's includes also that for closed QPT. The presence of a reservoir may also decimate criticality present in the closed case~\cite{joshi1}, which will turn out to be the case in some of the models we explore. Another possibility is that the Lindblad jump operators do not destroy criticality but alters the type of transition, {\it i.e.} the universality class is changed. 

Contrary to critical behaviour for closed systems, the Liouvillian $\hat{\mathcal L}_{\hat H,\,\hat L_i}$ can be non-analytical for finite systems in terms of so called {\it exceptional points}~\cite{ep}. However, for finite systems, single exceptional points does not normally give rise to PT's defined as non-analyticity in $\hat\rho_\mathrm{ss}$. Thus, the PT's we consider result from taking the thermodynamic limit also for the open models.

\section{Quantum optical models and the normal--superradiant phase transition}\label{sec3}
The NS PT derives from an instability of vacuum arising when the light-matter coupling is increased. For a critical coupling the photon mode and the excitations of the atoms become macroscopically populated also for the system ground state. The Dicke model has become the paradigmatic example for this type of transition and we start by revisiting the transition in this model, both for the closed case and the open one with photon losses present. We may note, however, that historically the transition was discussed earlier for the TC model~\cite{TCPT} than for the Dicke model~\cite{DickePT,DickePT2}.

\subsection{The Dicke model}
Two fairly recent experiments demonstrated the superradiant instability~\cite{dickeexp1,dickeexp2}, which surely boosted the general interest in NS PT's. The two experiments bear many similarities, but also differences. Common to both models are that they are driven by external lasers and that photon losses must be handled with care. The fact that the light-matter couplings are mediated by external lasers makes it possible to circumvent the {\it no-go theorem} that says that under a set of assumptions it is impossible to reach the transitions as long as the lower atomic electronic state is the atomic ground state~\cite{nogo}. It should, however, be noticed that the results of this theorem rely heavily on the assumptions made~\cite{vukics}. Since the PT instability appears also for zero temperature~\cite{zero}, the NS PT in the Dicke model has by most been termed a QPT. The idea of this Subsection is to revise this claim, or more precisely explore whether the transition is driven by quantum fluctuations at $T=0$. 

\begin{figure}[h]
\centerline{\includegraphics[width=11cm]{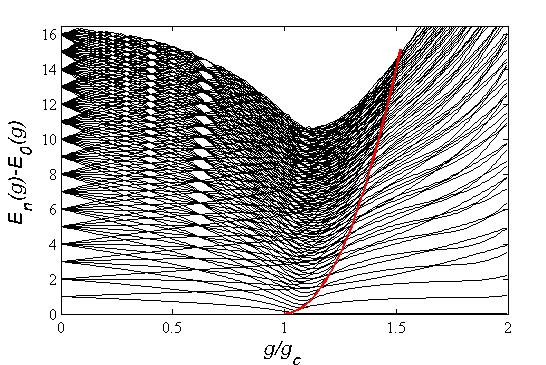}}
\caption{Spectrum of the resonant ($\omega=\Omega=1$) Dicke model. The number of atoms $N=80$ for which the closing of the gap at the critical point becomes clearly evident. At higher energies, states of different parity also merge to form a degenerate pair. The red thick line characterizes the boundary for this to happen, and it goes as $N(g-g_c)^2$. This scaling can be understood from the adiabatic potentials of (\ref{adham}); the depths of the potentials increase as $(g-g_c)^2$ meaning that given a certain energy more and more bound states appear as $g$ becomes larger. In the thermodynamic limit, the density of states $\nu(E)$ abruptly changes along the red line which represents an excited state phase transition~\cite{ept}.} \label{fig9}
\end{figure}

Let us use the traditional terminology, {\it i.e.} considering $N$ two-level `atoms' dipole-coupled to a common `photon mode'. As the total `spin' is conserved we introduce the collective spin operators
\begin{equation}
\hat S_\alpha=\sum_{i=1}^N\hat\sigma_\alpha^{(i)},\hspace{1cm}\alpha=x,\,y,\,z,
\end{equation}
where $\hat\sigma_\alpha^{(i)}$ is the $\alpha$-Pauli matrix acting on the two internal electronic states $|g\rangle$ and $|e\rangle$ for atom $i$. With $\omega$ the photon frequency and $\Omega$ the atomic transition frequency, the Dicke Hamiltonian reads ($\hbar=1$)
\begin{equation}
\hat H_\mathrm{D}=\omega\hat a^\dagger\hat a+\frac{\Omega}{2}\hat S_z+\frac{g}{\sqrt{N/2}}\left(\hat a^\dagger+\hat a\right)\hat S_x.
\end{equation}
The atom-light coupling $g$ has been rescaled by $1/\sqrt{N/2}$ in order to achieve non-trivial scaling in the thermodynamic limit $N\rightarrow\infty$; as written every term $\mathcal{O}(N)$ in the large $N$ limit. The photon creation/annihilation operators are $\hat a^\dagger/\hat a$, and note that the Hamiltonian can be composed in a `bare' $\hat H_0$ and `coupling' part $\hat H_1$, {\it i.e.} $\hat H_\mathrm{D}=\hat H_0+g\hat H_1$. For $g=0$ the ground state is the trivial vacuum $|\psi_0(0)\rangle=|0,-S\rangle$, where $|n,m\rangle$ denotes the state with $n$ photons and $\hat S_z|n,m\rangle=m|n,m\rangle$ and $S=N/2$ is the total spin (we consider the spin sector with maximum spin as it incorporates the lowest energy state). Thus, all atoms occupy their lower electronic state $|g\rangle$. The Hamiltonian is symmetric under the parity operator $\hat\Pi=\exp\left[-i\pi\left(\hat a^\dagger\hat a+\hat S_z/2\right)\right]$ and it thereby supports a $\mathbb{Z}_2$ symmetry. It is this symmetry that is broken in the superradiant phase, and the corresponding critical coupling $g_c=\sqrt{\omega\Omega}/2$. The spectrum for the lowest 150 eigenvalues of the Dicke model is shown in Fig.~\ref{fig9}. Here $N=80$ and the gap closing at the critical point is clearly visible; in the superradiant phase we see the merging of the two parity eigenstates. Not evident from the figure, but the larger $N$, the more clear is it that a step-like behaviour develops in $\partial_g^2E_0(g)$ around $g=g_c$. The thick red line marks a discontinuity in the density of states $\nu(E)$ in the limit $N\rightarrow\infty$. It scales as $\sim N(g-g_c)^2$ which can be understood from the mean-field treatment outlined below. 

We may gain further insight by rewriting the Hamiltonian in its field quadratures
\begin{equation}
\hat x=\frac{1}{\sqrt{2}}\left(\hat a^\dagger+\hat a\right),\hspace{1cm} \hat p=\frac{i}{\sqrt{2}}\left(\hat a^\dagger-\hat a\right),
\end{equation}
{\it i.e.}
\begin{equation}
\hat H_\mathrm{D}=\omega\frac{\hat p^2}{2}+V_\mathrm{eff}(\hat x)=\omega\frac{\hat p^2}{2}+\frac{\Omega}{2}\hat S_z+\omega\frac{\hat x^2}{2}+\frac{2g}{\sqrt{N}}\hat x\hat S_x.
\end{equation}
Thus, the Dicke model can be viewed as a particle of mass $\omega^{-1}$ in a matrix-valued potential $V_\mathrm{eff}(\hat x)$. The action of the parity transformation is $\hat\Pi\hat x\hat\Pi^{-1}=-\hat x$,  $\hat\Pi\hat p\hat\Pi^{-1}=-\hat p$, and $\hat\Pi\hat S_x\hat\Pi^{-1}=-\hat S_x$, while leaving $\hat S_z$ unaltered. We transform the Hamiltonian to the {\it adiabatic representation} by the $\hat U$ defined as $\hat U(\hat x)V_\mathrm{eff}(\hat x)\hat U^{-1}(\hat x)=D(\hat x)=\omega\frac{\hat x^2}{2}+\hat S_z\sqrt{\frac{\Omega^2}{4}+\frac{4g^2}{N}\hat x^2}$~\cite{boa0};
\begin{equation}\label{hamtr}
\hat H_\mathrm{D}'=\omega\frac{\left(\hat p^2-i\hat U(\hat x)\partial_{\hat x}\hat U^{-1}(\hat x)\right)^2}{2}+D(\hat x).
\end{equation}
The {\it Born-Oppenheimer approximation}~\cite{boa} consists in neglecting the `gauge potential' 
\begin{equation}
\hat A(\hat x)=-i\hat U(\hat x)\partial_{\hat x}\hat U^{-1}(\hat x)=\frac{\Omega^2N}{\Omega^2N+16g^2\hat x^2}\hat S_y
\end{equation}
from (\ref{hamtr}) such that $\hat H_D'$ is diagonal in the spin subspace. Naturally, $\hat A(\hat x)$ is a non-diagonal $(2S+1)$-dimensional matrix which causes couplings between the {\it adiabatic states}. It should be noted that this approximation becomes exact if $[\hat x,\hat p]=0$ and it should therefore be seen as a mean-field approximation -- we neglect quantum fluctuations. In the thermodynamic limit the approximation is indeed exact, see Refs.~\cite{DickePT2,Dickethermo}. One can show the exactness very intuitively. We define the `boson' operators $\hat b=\hat a/\sqrt{N}$ and $\hat b^\dagger=\hat a^\dagger/\sqrt{N}$ such that we can write the full Hamiltonian as a sum of $N$ quantum Rabi model (see Subsec.~\ref{subsecRabi} below) with the $\hat b$-operators. But this means that in the thermodynamic limit $\left[\hat b,\hat b^\dagger\right]\rightarrow0$ and we can ignore fluctuations in the boson field. It is then straightforward to calculate, for example, the partition function using the {\it transfer matrix method}~\cite{DickePT2}. Neglecting quantum fluctuations is equivalent to considering the energy $E[\alpha]=\langle\alpha|\hat H_\mathrm{D}|\alpha\rangle$ where we introduced the coherent state, $\hat a|\alpha\rangle=\alpha|\alpha\rangle$, with complex amplitude $\alpha$. Fixed points of this functional give the classical or mean-field solutions. Summing up, in the thermodynamic limit we can reduce the analysis of the adiabatic Hamiltonian to the lowest potential surface of~(\ref{hamtr}), {\it i.e.}~\cite{dickeboa}
\begin{equation}\label{adham}
\hat H_\mathrm{D}^{(ad)}=\omega\frac{\hat p^2}{2}+\omega\frac{\hat x^2}{2}-\frac{N}{2}\sqrt{\frac{\Omega^2}{4}+\frac{4g^2}{N}\hat x^2}.
\end{equation}
For $g>g_c=\sqrt{\omega\Omega}/2$ the potential builds up a double-well structure representing the transition from the normal to the superradiant phase. The value of the coherent state amplitude, given by the minima of the energy surface, becomes 
\begin{equation}\label{amp}
\alpha=(x+ip)/\sqrt{2}=\pm\sqrt{N}\left(\frac{g^2}{\omega^2}-\frac{\Omega^2}{16g^2}\right)^{1/2}.
\end{equation} 
In the symmetry broken phase when a single minimum of the double-well is chosen, the phase of the field is 0 or $\pi$. This manifestation of the symmetry breaking was experimentally verified in Ref.~\cite{phaseDicke}. Note that the average photon number $n=\alpha^*\alpha\sim N$ and the corresponding critical exponent $\beta=1$~\cite{critDicke}.
 
Already the previous paragraph tells us that the NS PT in the Dicke model cannot be a QPT in the sense that it is driven by quantum fluctuations. Nevertheless, we point out that the ground state energy $E_0(g)$ shows a non-analytic behaviour at the critical point. Let us elaborate further on the fact that quantum fluctuations can be disregarded in the thermodynamic limit as it seems to be not generally accepted. In Sec.~\ref{sec2} we gave some general properties of PT's and QPT's, for example the divergence of a length scale upon approaching the critical point. In the Dicke model there is no such divergence since we do not have any inherent length scale to start with. The model is of the type {\it fully connected} meaning that every atom interacts with every other atom~\cite{fc}. Here, the photon mode mediates the interaction among all the atoms and in this non-relativistic framework the effective coupling is non-local. In this respect the Dicke model differ from the paradigmatic models (like the Ising model in a transverse field) mostly encountered when discussing QPT's. For a single atom/spin it is as if it sees an infinite number of neighbours (in the thermodynamic limit). But the essence of the Mermin-Wagner theorem is that in higher dimensions quantum fluctuations are of less importance, and in particular the mean-field approximation should, in fact, become exact in the infinite dimensional limit~\cite{inf}. Indeed, the Dicke PT is a so called {\it mean-field phase transition} where the mean-field critical exponents are exact~\cite{critDicke}. This is another demonstration that the transition is not a proper QPT. As an additional motivation for this we may consider the range of fluctuations in the finite $N$ case. For the example of the Ising model in Sec.~\ref{sec2} we have that the `total amount of fluctuations' grows as $\sim N$ with the system size. This means that the fluctuations are non-negligible relative to the system size in the thermodynamic limit. In the Dicke model, however, the fluctuations of both the macroscopic spin and the field scale as $\sim\sqrt{N}$, and it follows that the fluctuations are in this case negligible relative to other scales in the thermodynamic limit. This is not surprising, after all the thermodynamic limit of the Dicke model is the typical classical limit for the spin. The different scaling between the Dicke and the Ising model derives from the symmetries of the Dicke model where the total spin is preserved such that we can work with the collective spin operators and reduce the space to a single spin sector. In the Ising model, on the contrary, the thermodynamic limit does not represent such a classical limit -- the model remains highly quantum. In particular, the Dicke model belongs to the same class as the infinite range transverse field Ising model which is identified as a special kind of the {\it Lipkin-Meshkov-Glick model}~\cite{lmg}. This also becomes clear in the {\it Holstein-Primakoff representation} where the boson mode can be mapped into a spin~\cite{brandes}. Alternatively, starting from the Dicke model and eliminating the photon field one ends up with the Lipkin-Meshkov-Glick model~\cite{lmg2}. It is interesting, and surprising though, that in the limit $N\rightarrow\infty$ the ground state of the Dicke model shows persistent entanglement in the spin degree-of-freedom~\cite{vidal}.

Experimentally it is most relevant to understand how the Dicke PT is affected by especially photon losses. Thus we consider the steady state of the master equation~\cite{scully}
\begin{equation}\label{lindbladDicke}
\partial_t\hat\rho=i\left[\hat\rho,\hat H_\mathrm{D}\right]+\kappa\left(2\hat a\hat\rho\hat a^\dagger-\hat a^\dagger\hat a\hat\rho-\hat\rho\hat a^\dagger\hat a\right)
\end{equation}  
describing the coupling of the system to a Markovian zero temperature bath. It is clear that we cannot simply generalize the Born-Oppenheimer mean-field theory that identifies adiabatic potential surfaces to this open problem. However, under the assumption of a large spin (thermodynamic limit), it is justifyable to apply the factorization that is related to letting operators commute and then write down the equations of motions $\partial_t\langle\hat A\rangle=\partial_t\mathrm{Tr}\left[\hat A\hat \rho\right]$. In the quadrature representation we find~\cite{carmichael}
\begin{equation}
\begin{array}{l}
\displaystyle{\dot x=\omega p-\kappa x,}\\
\displaystyle{\dot p=-\omega x-\frac{2gS_x}{\sqrt{N}}-\kappa p,}\\
\displaystyle{\dot S_x=-\frac{\Omega}{2}S_y,}\\
\displaystyle{\dot S_y=\frac{\Omega}{2}S_x-\frac{2g}{\sqrt{N}}xS_z},\\
\displaystyle{\dot S_z=\frac{2g}{\sqrt{N}}xS_y,}
\end{array}
\end{equation}
where we have written the expectations $A=\langle\hat A\rangle$ and the dot represents time derivative. The critical coupling is shifted by the presence of photon losses according to
\begin{equation}\label{dickecrit} 
g_c=\frac{1}{2}\sqrt{\frac{\Omega}{\omega}\left(\kappa^2+\omega^2\right)}.
\end{equation} 
There is one trivial steady state of the equations of motion corresponding to the normal phase and which is stable whenever $g<g_c$, {\it i.e.}
\begin{equation}
(x,p,S_x,S_y,S_z)_\mathrm{ss}=(0,0,0,0,-N/2)
\end{equation}
Above the critical coupling this solution becomes unstable and two new symmetric stable solutions emerge~\cite{carmichael},
\begin{equation}\label{dickess}
\begin{array}{l}
\displaystyle{(x,p)_\mathrm{ss}=\pm(\omega,\kappa)\frac{\sqrt{2N}g}{\omega^2+\kappa^2}\sqrt{1-\frac{g_c^4}{g^4}}},\\ \\
\displaystyle{(S_x,S_y,S_z)_\mathrm{ss}=\frac{N}{2}(\pm\sqrt{1-\frac{g_c^4}{g^4}},0,-\frac{g_c^2}{g^2})}.
\end{array}
\end{equation} 
Note that whenever $\kappa\neq0$ the $p$-quadrature is non-zero in the superradiant phase, which is a result of the Lamb shift arising when the system couples to a reservoir. The NS PT manifests in terms of a {\it pitchfork bifurcation} and in particular, the bifurcation survives even at $\kappa\neq0$. However, it has been shown that some critical exponents are no longer the same when photon losses are taken into account~\cite{critDicke,critDicke2}.

\subsection{The Tavis-Cummings model}
\begin{figure}[h]
\centerline{\includegraphics[width=11cm]{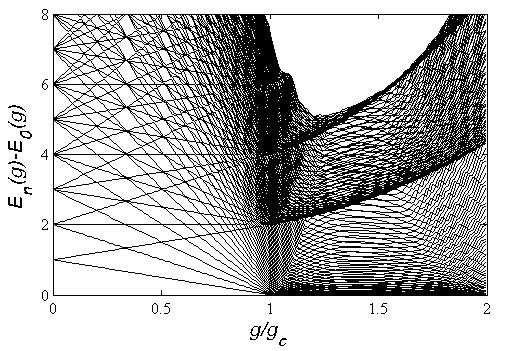}}
\caption{TC spectrum (150 lowest energies) for $N=80$ atoms. By considering the resonance situation $\omega=\Omega$ the spectrum for $g=0$ is equidistant in integer steps. For each integer the degeneracy at $g=0$ increases by one corresponding to the Dicke states belonging to different $\mathcal{K}$ sectors. Naturally, the energies of different $\mathcal{K}$ sectors do cross, meaning that the TC PT at $g=g_c$ cannot be driven by quantum fluctuations. In the superradiant phase we see also a steplike behaviour in the density of states $\nu(E)$ where the steps are related to $\hat S_z$ sectors. The sharp 'scars' caused by these steps define excited state PT's. In the thermodynamic limit, $N\rightarrow\infty$, continuous bands are formed starting at the steps. This is a result of the broken continuous symmetry that implies a Goldstone mode.} \label{fig10}
\end{figure}

The TC model derives from the Dicke one by applying the RWA~\cite{rwa},
\begin{equation}
\hat H_\mathrm{TC}=\omega\hat a^\dagger\hat a+\frac{\Omega}{2}\hat S_z+\frac{g}{\sqrt{N/2}}\left(\hat a^\dagger\hat S^-+\hat S^+\hat a\right),
\end{equation}
with $\hat S^\pm$ the spin ladder operators. The total number of excitations $\hat K=\hat a^\dagger\hat a+\hat S_z/2$ is preserved, {\it i.e.} within the RWA the discrete $\mathbb{Z}_2$ parity symmetry of the Dicke model turns into a continuous $U(1)$ symmetry. The higher symmetry implies that the TC model is integrable contrary to the Dicke model for general $N$. A result of the integrability is seen in the spectrum as true crossings between energy eigenstates belonging to different $\mathcal K$-sectors. The 150 lowest eigenvalues as a function of the coupling strength are displayed in Fig.~\ref{fig10}. As anticipated, for small couplings $g<g_c/10$, when the RWA is applicable, the Dicke and TC models show the same structure. For $g_c/10$ to $\sim g_c/2$ the difference between the models is evident; the {\it Bloch-Siegert shift} is of order several percent. For even larger couplings the two spectra are no longer qualitatively similar, and closer to the critical point the symmetries manifest in terms of degeneracies. In the thermodynamic limit the spectrum is gapless in the superradiant phase representing the Goldstone mode. More precisely, when $N\rightarrow\infty$ infinitely many eigenvalues intersect at $g=g_c$, one from every excitation sector $\hat K$. For finite $N$, the crossings are split along a line: First the $K=-N/2$ cross with the $K=-N/2+1$, then the $K=-N/2+1$ with the  $K=-N/2+2$, and so on. In the Appendix we discuss some properties of the $X\!X$ model and especially for the finite $X\!X$ model the ground state energy $E_0(g)$ also exhibits a series of true crossings, and in the thermodynamic limit these all coalesce at a single critical coupling $g_c=1/2$~\cite{xx2}. In this respect, like for the TC model, the ground state energy displays a non-analyticity.  In both models, the TC and the $X\!X$, the true crossings stem from particle conservation, bare excitations in the TC model and number of fermions in the $X\!X$ model. As a result, there are in a strict sense {\bf no} quantum fluctuations driving the transition. In particular, the TC Hamiltonian can be decomposed on block form with every block characterized by a $K$ quantum number. Nothing couples the blocks and we cannot have direct transitions between them. As an example, assume $g=0$ and the system is in its ground state, the vacuum. If it somehow would be possible to adiabatically ramp up $g$ across the critical point, the system would stay in the vacuum even though it is no longer the true ground state. Not surprising, it seems impossible to couple states of different $\mathcal K$ sectors without breaking the $U(1)$ symmetry. Photon losses would indeed allow for transitions between the $\mathcal K$ sectors, but the question is whether such a system can be considered to support a continuous symmetry.

As for the Dicke model, let us apply the Born-Oppenheimer approximation also for the TC model. Thus, we diagonalize the spin part as if the boson operators would be $c$-numbers, {\it i.e.} commute. In the quadrature representation the TC model is
\begin{equation}\label{tc2}
\hat H_\mathrm{TC}=\omega\frac{\hat p^2}{2}+\frac{\Omega}{2}\hat S_z+\omega\frac{\hat x^2}{2}+\frac{2g}{\sqrt{N}}\left(\hat x\hat S_x-\hat p\hat S_y\right),
\end{equation}
and the corresponding adiabatic Hamiltonian~\cite{jonasberry}
\begin{equation}\label{adham2}
\hat H_\mathrm{TC}^{(ad)}=\omega\frac{\hat p^2}{2}+\omega\frac{\hat x^2}{2}-\hat S_z\sqrt{\frac{\Omega^2}{4}+\frac{4g^2}{N}\left(\hat x^2+\hat p^2\right)}.
\end{equation}
With the same comparison as for the Dicke vs Ising models, we can compare the TC with the $X\!X$ model presented in the Appendix. The quadratures $\hat x$ and $\hat p$ are then identified as the polarizations in the $x$- and $y$-directions respectively. In the superradiant phase the lowest adiabatic energy surface, $\hat S_z=-N/2$, builds up a sombrero shape in the $xp$-plane, which reflects the $U(1)$ symmetry. At the mean-field level, when $[\hat x,\hat p]=0$, any point $(x,p)_\mathrm{ss}$ along the potential minimum of the sombrero is a ground state solution. Schematically, translations along the minimum do not cost energy and the spectrum is hence gapless. 

Something surprising occurs as we include photon losses, {\it i.e.} considering the Lindblad master equation~(\ref{lindbladDicke}) for the TC model. The mean-field equations-of-motion become
\begin{equation}
\begin{array}{l}
\displaystyle{\dot x=\omega p-\frac{2g}{\sqrt{N}}S_y-\kappa x,}\\
\displaystyle{\dot p=-\omega x-\frac{2g}{\sqrt{N}}S_x-\kappa p,}\\
\displaystyle{\dot S_x=-\frac{\Omega}{2}S_y+\frac{2g}{\sqrt{N}}pS_z,}\\
\displaystyle{\dot S_y=\frac{\Omega}{2}S_x+\frac{2g}{\sqrt{N}}xS_z},\\
\displaystyle{\dot S_z=\frac{2g}{\sqrt{N}}\left(xS_y+pS_x\right).}
\end{array}
\end{equation}
For $\kappa=0$, the steady state solutions agree with the predictions of the lowest Born-Oppenheimer energy surface~(\ref{adham2}). However, when we turn on $\kappa$ the steady-state solution corresponding to a superradiant phase vanishes. Thus, the decay rate of these states towards the unique trivial steady state $|0,-N/2\rangle$ is non-zero. In other words, contrary to the Dicke model where photon losses only alters the criticality of the closed model, for the open TC model they completely decimate the criticality. We have so far not shown that the mean-field and full quantum results agree for the TC model in the thermodynamic limit and even less for the open TC model. Thus, it is not clear whether vacuum is also the unique steady state of the full Lindblad master equation. As we will see numerically, it indeed is and it can furthermore be shown in a very intuitive way.

\begin{figure}[h]
\centerline{\includegraphics[width=11cm]{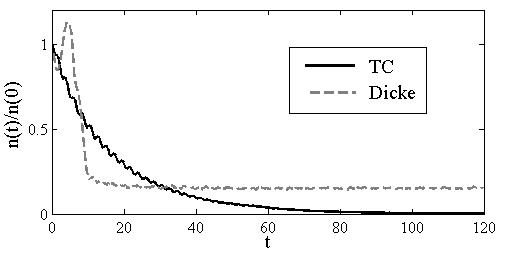}}
\caption{Evolution of the scaled photon number for the TC (solid black line) and Dicke models (dashed grey line). The initial state is the corresponding ground state of the Hamiltonian, which for the present parameters has an average photon number $\sim17$ for the TC model and $\sim88$ for the Dicke model. The larger photon number for the Dicke model derives from the fact that the critical point is half that of the TC model and thereby the system is deeper in the superradiant phase. The decay is to a good approximation exponential for the TC model in agreement with the mean-field predictions. As a comparison, the Dicke model which supports a non-trivial steady state, does not display the same exponential decay. Note, however, that the average photon number is lowered by the photon dissipation which is in agreement with (\ref{dickecrit}). The number of atoms $N=40$, and the remaining dimensionless parameters are $\kappa=0.1$, $g=1.5$, and $\omega=\Omega=1$. } \label{fig11}
\end{figure}

Symmetries for Lindblad master equations are defined as~\cite{lindsym}
\begin{equation}
\hat\mathcal{L}_{\hat H,\,\hat a}\left[\hat\rho\right]=\hat\mathcal{L}_{\hat U\hat H\hat U^{-1},\,\hat U\hat a\hat U^{-1}}\left[\hat\rho\right],
\end{equation}
for some unitary $\hat U(\varphi)$. It is clear that this is satisfied for the TC model with photon losses and $\hat U(\varphi)=\exp\left(i\hat K\varphi\right)$. Thus, the continuous symmetry of the TC model is not lifted by photon losses. At first this seems strange, but for Lindblad master equations -- generators of non-unitary time-evolution, there is no correspondence of Noether's theorem, {\it i.e.} a symmetry does not necessarily imply a conserved quantity~\cite{lindsym}. The present situation is exactly such a case; even though the Lindblad master equation is symmetric under $\hat U$, an initial state $\hat\rho_{K}(0)$ with a well defined excitation $K$ will not remain in its initial $\mathcal K$ sector. Indeed, the unique steady state is the vacuum such that any initial state will eventually end up in the $K=-N/2$ sector. This has been numerically confirmed in Fig.~\ref{fig11} showing the results from a {\it quantum jump} simulation~\cite{jump}. The plot shows the evolution of the scaled photon number $n(t)/n(0)$ with $n(t)=\langle\hat a^\dagger\hat a\rangle_t$ for an initial state being the ground state of the Hamiltonian and $N=40$ atoms. Apart from the small wiggles, the decay is exponential with a decay rate somewhat larger than $\kappa^{-1}$. The fact that the steady state $\hat\rho_\mathrm{ss}=|0,-N/2\rangle\langle0,-N/2|$, independent of parameters, implies that the open TC model is {\bf not} critical. This is in contrast to the Dicke model which indeed is critical also in the presence of photon losses. In the same figure we demonstrate this by showing the same quantity but for the Dicke model, and it is evident that the steady state is not the trivial vacuum. We should, however, note that to a first approximation $\kappa^{-1}$ sets the time scale for reaching the steady state, and for shorter scales one may envision (dynamical) critical behaviour also in the TC model. That is, if there is an internal (relaxation) time-scale short in comparison to $\kappa^{-1}$ it might happen that the system seems to occupy a superradiant state, however unstable. Nevertheless, in a strict sense the open TC model is not critical.

There is a simple explanation why the Dicke model reaches a non-trivial state, while the TC model does not. Let us decompose the system state $\hat\rho$ in blocks belonging to the different $\mathcal K$ sectors;
\begin{equation}
\hat\rho=\left[
\begin{array}{cccc}
\rho_{00} & \rho_{01} & \rho_{02} & \cdots\\
\rho_{10} & \rho_{11} & \rho_{12} & \cdots\\
\rho_{20} & \rho_{21} & \rho_{22} & \cdots\\
\vdots & \vdots & \vdots & \ddots
\end{array}\right].
\end{equation}
Here, $\rho_{ii}$ is the matrix with elements of the density operator that belong to the $\mathcal K$ sector with $i$ excitations, and the $\rho_{ij}$'s are the `coherences' between the $\mathcal K$ sectors with $i$ and $j$ excitations. For the closed Dicke model, $\hat K$ is not conserved and time-evolution may cause population transfer between the sectors; $\rho_{i+1i+1}\leftrightarrow\rho_{ii}$. For the closed TC model such transfer is naturally prohibited by symmetry. Nevertheless, photon losses allows for the irreversible transfer $\rho_{i+1i+1}\rightarrow\rho_{ii}$. As a result, dynamically population can shift upward along the diagonal in the open TC model, but not downward, and hence the steady state will necessarily be $\hat\rho_\mathrm{ss}=\rho_{00}$. In the Dicke model, on the contrary, the counter rotating terms take population downward along the diagonal and the steady state is given by the detailed balance between the processes -- upward vs. downward.

\subsection{The quantum Rabi model}\label{subsecRabi}
At first sight, criticality in the quantum Rabi model~\cite{rabi}
\begin{equation}
\hat H_\mathrm{R}=\omega\left(\frac{\hat p^2}{2}+\frac{\hat x^2}{2}\right)+\frac{\Omega}{2}\hat\sigma_z+\sqrt{2}g\hat x\hat\sigma_x
\end{equation}
seems impossible since there is no clear thermodynamic limit. For the Dicke model, in the normal phase a single {\it bare state} $|0,-N/2\rangle$ is populated, while in the superradiant phase the number of populated bare states scales as $\sim N$. Put in other words, in the superradiant phase the photon mode becomes macroscopically populated when we take the thermodynamic limit, {\it i.e.} letting the spin become `classical'. We may flip the coin and consider the `classical limit' of the harmonic oscillator instead which would be represented by $\omega/\Omega\rightarrow0$~\cite{claslim}. In a set of independent papers it was realized that in such a limit critical-like behaviour and a symmetry breaking are also possible in the Rabi model~\cite{rabipt1,rabipt2,plenio1}. 

\begin{figure}[h]
\centerline{\includegraphics[width=11cm]{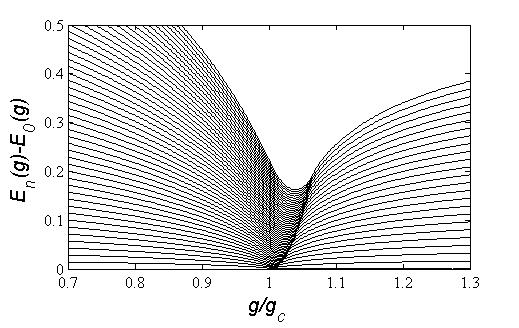}}
\caption{Spectrum of the Rabi model $\hat H_\mathrm{R}^{(C)}$ for the 50 lowest energies, and $C=2500$ and non-scaled frequencies $\omega=\Omega=1$. The closing of the energy gap at the critical point is visible as is the line marking the excited state PT~\cite{plenio3}.} \label{fig12}
\end{figure}

As $\omega\rightarrow0$ the discreteness of the harmonic oscillator spectrum is suppressed and it resembles more the classical one. Naturally, as the energy scale of the atom is much larger one can envision that the atom can cause a large number of excitations of the oscillator. To hold the critical coupling fixed we introduce a new variable that scales the frequencies
\begin{equation}
\hat H_\mathrm{R}^{(C)}=\frac{\omega}{\sqrt{C}}\left(\frac{\hat p^2}{2}+\frac{\hat x^2}{2}\right)+\frac{\sqrt{C}\Omega}{2}\hat\sigma_z+\sqrt{2}g\hat x\hat\sigma_x.
\end{equation}
With the scaling $\hat\sigma_\alpha\sim1$, $\hat x\sim\hat p\sim \sqrt{C}$ we have that every term in the rescaled Hamiltonian goes as $\sim\sqrt{C}$. The new thermodynamic limit consists in $C\rightarrow\infty$, and we expect a non-analyticity at $g_c=\sqrt{\omega\Omega}/2$ as for the Dicke model. It should be appreciated that the two thermodynamic limits represent the classical limits of the spin (Dicke) and the oscillator (Rabi).

In the Born-Oppenheimer approximation, which is expected to be justified in the thermodynamic limit of very large $\sqrt{C}\Omega$, we recover the effective lower adiabatic potential
\begin{equation}\label{adpot2}
V^{(ad)}(\hat x)=\frac{\omega}{\sqrt{C}}\frac{\hat x^2}{2}-\sqrt{\frac{C\Omega^2}{4}+2g^2\hat x^2}.
\end{equation} 
In the superradiant phase, this mean-field approach predicts a field quadrature amplitude
\begin{equation}
x=\sqrt{2C}\left(\frac{g^2}{\omega^2}-\frac{\Omega^2}{16g^2}\right)^{1/2}=\frac{\sqrt{2C}}{\omega g}\left(g^4-g_c^4\right)^{1/2}.
\end{equation}
Comparing to the corresponding expression (\ref{amp}) for the Dicke model we see that $C$ serves the same role as $N$. Following the same procedure as for the Dicke model one also finds the mean-field steady states for the atom, $\left(\hat\sigma_x,\,\hat\sigma_y,\,\hat\sigma_z\right)_\mathrm{ss}$, identical to those of the Dicke model, Eq.~(\ref{dickess}), with $N=1$. While the mean-field approaches of the Dicke vs. the Rabi models show great resemblances in their respective thermodynamic limits, it should be remembered that criticality in the Rabi model occurs only in the extreme dispersive regime. It has been argued that this regime could be reached within trapped ion systems where the photon mode is replaced by the harmonic motion of the ion and the internal states are coupled to the vibrational ones via spatially dependent lasers~\cite{plenio2}. Furthermore, even though the two models are identical at the mean-field level, the quantum models are qualitatively different. For example, the Dicke model is quantum chaotic showing {\it level repulsion}~\cite{brandes}, while the Rabi model is quantum integrable with no pronounced level repulsion~\cite{braak}. 

We argued that the Dicke PT is not the typical text-book example of a QPT with a diverging length scale at the critical point and a thermodynamic limit consisting in letting the degrees-of-freedom become infinite. Instead, the Dicke PT is a mean-field transition for which quantum fluctuations are vanishingly small (relative to the system size) in the thermodynamic limit. Now, for the Dicke model the thermodynamic limit consists in taking the classical limit of the spin, while in the Rabi model criticality appears when taking the classical limit of the boson mode instead. Thus, the spin is still very `quantum' and one may ask whether the transition can be driven by quantum fluctuations or not. A qualified guess is that this cannot be the case since the fluctuations of the spin will be negligibly small in comparison to those of the boson. Regarding critical exponents we note, for example, that for the Rabi model the mean-field approach predicts that the photon number in the vicinity of the critical point in the superradiant phase 
\begin{equation}
n\propto|\lambda-\lambda_c|^\mu
\end{equation}
with $\lambda=g^4$, $\lambda_c=g_c^4$, and the critical exponent $\mu=1$. Numerical exact diagonalization of the Rabi model for a scaling coefficient $C$ as large as $10\,000$ shows that also the full quantum model shares the same critical exponent. Even though it does not make much sense to talk about the Rabi model as `fully connected' like the Dicke one, it is suggested that the mean-field predictions are exact in the thermodynamic limit. Indeed, similar arguments could also be used as for the Dicke model, {\it i.e} introducing the scaled boson operators $\hat b=\hat a/\sqrt{C}$ and $\hat b^\dagger=\hat a^\dagger/\sqrt{C}$ and proving, using Wick's theorem, that quantum fluctuations can be omitted as $C\rightarrow\infty$~\cite{Dickethermo}. Hence, the Rabi PT is also of the mean-field type just like the Dicke PT.

We end this section by showing the spectrum of the Rabi model for a large $C$, Fig.~\ref{fig12}. This plot cannot be directly compared to Fig.~\ref{fig9} showing the spectrum of the Dicke model since that one is for the resonant case $\omega=\Omega$, while the present one of the critical Rabi model is for the extremely detuned situation, $\sqrt{C}\Omega\gg\omega/\sqrt{C}$. Still we see a parabolic line representing the excited state PT~\cite{plenio3}, and the gap closing due to the $\mathbb Z_2$ symmetry breaking is evident.

\subsection{The Jaynes-Cummings model}
\begin{figure}[h]
\centerline{\includegraphics[width=11cm]{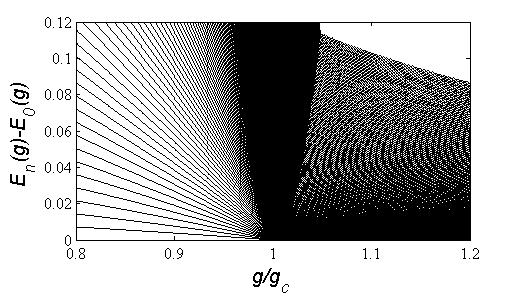}}
\caption{Spectrum (lowest 250 energies) for the JC model with $C=2500$ ($\omega=\Omega=1$). Due to the continuous symmetry, in the limit $C\rightarrow\infty$ an energy continuum builds up in the superradiant phase $g>g_c$. As for the other models, a clear `scar' in the spectrum is visible which represents an excited state PT. } \label{fig13}
\end{figure}

Hwang and Plenio considered the bosonic classical limit also for the JC model~\cite{plenio4} and, expectedly, it shows great resemblances with the criticality in the TC model. The analysis goes as in the previous subsection for the quantum Rabi model but after application of the RWA. Thus, we consider the scaled JC Hamiltonian
\begin{equation}
\hat H_\mathrm{JC}^{(C)}=\frac{\omega}{\sqrt{C}}\hat a^\dagger\hat a+\frac{\sqrt{C}\Omega}{2}\hat\sigma_z+g\left(\hat a^\dagger\hat \sigma^-+\hat \sigma^+\hat a\right).
\end{equation}
The 250 lowest energies of $\hat H_\mathrm{JC}^{(C)}$ are displayed in Fig.~\ref{fig13}. It is clear how a large number of energies combine at the critical point to form a continuous band in the thermodynamic limit $C\rightarrow\infty$. Again, signatures of an excited state PT are also evident.

Like for the TC, $\hat K=\hat a^\dagger\hat a+\hat\sigma_z/2$ is preserved for the JC model. This implies that all the energies to the left of Fig.~\ref{fig13} belong to different $\mathcal K$ sectors and as they cross there is no coupling between the corresponding states. Just like the TC model, when $C\rightarrow\infty$ all the infinite number of crossings of the ground state energy merge at $g_c$ to produce a single point of non-analyticity of $\partial_g^2E_0(g)$ as for a continuous PT. However, there is no instability of the vacuum state due to quantum fluctuations.

Furthermore, the same arguments applies to the open JC model as for the open TC model, namely that in the presence of photon losses the unique steady state is the trivial vacuum; $\hat\rho_\mathrm{ss}=|0,-1/2\rangle\langle0,-1/2|$. Consequently the open JC model is not critical. One may naturally argue that there are other open JC models (and open TC models) with different Lindblad jump operators. The demonstration given above that the steady state is the vacuum for photon losses can be readily applied also to atomic spontaneous emission with $\hat L=\hat\sigma^-$ with the same conclusion; excitations are dissipated but nothing counteracts this dissipation. However, for the JC model spontaneous emission causes a much slower relaxation of the system to its steady state than photon losses  do since the atom cannot be multiply excited like the photon mode. Other Lindblad jump operators, not representing dissipation, for example dephasing $\hat L=\hat\sigma_z$, do give non-trivial steady states for the JC and TC models, but those are not the ones we are concerned with here.

\section{Concluding remarks}
In this paper we discussed some aspects of the NS PT that appears in the paradigmatic quantum optical models; Dicke, Rabi, TC, and JC. One ambition has been to put the discussion on a general setting in terms of QPT's. It has been a consensus that the NS PT's in these models are indeed QPT's. In the literature you typically hear that a PT occurring at zero temperature is a QPT. As we have argued, the meaning of such a statement needs to be clarified and especially a proper definition of a QPT is essential. Maybe the most common description is that a QPT is a transition that appears at zero degrees, $T=0$, and is possible due to quantum fluctuations. Taking this as a definition, one objection of the present paper has been to show that the transitions in neither the Dicke nor the TC model are QPT's. We have used several arguments supporting this for the Dicke model; it is a mean-field transition, the thermodynamic limit is a classical limit, fluctuations are vanishingly small in the classical limit, which can also be shown explicitly with the help of Wick's theorem. For the TC model, the absence of quantum fluctuations driving the PT is evident since the transition occurs between different symmetry sectors. However, one may alternatively define a QPT as a non-analyticity of the ground state, and not care whether it arises due to competing non-commuting terms of the Hamiltonian or due to crossings between different symmetry sectors. In this respect, both the Dicke and TC models are critical supporting continuous QPT's. 
  
%\begin{center}
\begin{table}[h]
 \begin{tabular}{| p{2.8cm} | l | l | p{3.1cm} | p{2.8cm} | p{2.8cm} | }
  \hline			
  Model & Critical & PT type & Critical exponent $\langle \hat n\rangle\propto|\lambda-\lambda_c|^\mu$ & Non-analyticity in system state & Quantum fluctuations \\
  \hline 
  \hline
  Dicke \& Rabi & Yes & Mean-field & $\mu=1$ & Yes (ground state)& Vanishes in thermo. limit \\
  \hline
  TC \& JC & Yes & Mean-field & $\mu=1$ & Yes (ground state) & No \\
  \hline
  Open 
  Dicke \& Rabi & Yes & Mean-field & $\mu=1$ & Yes (steady state) & Vanishes in thermo. limit\\
  \hline  
   Open TC \& JC & No & None & -- & No & No\\
  \hline 
 \end{tabular}
 \caption{Summary of the properties of the various models considered in this work. By ``open'' we here mean Markovian photon losses. All four closed models are quantum critical in the sense that the second derivative of the ground state energy, $\partial_g^2E_0(g)$, is discontinuous for $g=g_c$, but they are not quantum critical in the sense that there are no quantum fluctuations that destabilize the symmetric phase. While the critical exponent $\mu$ is identical for both the open and closed Dicke and Rabi models, there are other exponents that are changed due to the openness, for example how the photon fluctuations scale~\cite{critDicke}. Importantly, the open TC and JC models are not critical at all.}
 \end{table}\label{tab1}
 %\end{center}

We have compared and contrasted the Dicke and TC PT's with those of the transverse field Ising and $X\!X$ Heisenberg models in one dimension. There are similarities between these two pairs of models, and in particular the Dicke and TC models can qualitatively be seen as the infinite range counterparts of these one dimensional spin chains. This, again, shows that quantum fluctuations are not a driving mechanism of the transitions -- the critical behaviours of the mean-field models agree with the exact quantum results. This is opposite to the Ising model where the mean-field and exact quantum results predict different critical exponents. The TC and $X\!X$ models share some qualitative features of their respective spectra that derive from a particle conservation symmetry. For a finite system and for increasing coupling the ground state goes through an array of crossings between different particle $\mathcal K$-sectors where for every crossing the particle number increases by one unit. Naturally, there are no quantum fluctuations responsible for causing transitions between the different sectors. The ground state energy has a discontinuous first derivative $\partial_g E_0(g)$ at every crossing which is characteristic for a first order PT, but these non-analyticities do not originate from taking the thermodynamic limit and should therefore not be consider proper PT's. The interesting feature of these models occurs in the thermodynamic limit as these crossings merge into one critical point and $\partial_g^2 E_0(g)$ is instead discontinuous. This is the emergence of criticality in both these models. There is an important difference between the TC and the $X\!X$ model; the one dimensional $X\!X$ model with nearest neighbour interaction is prohibited to build up long-range order by the Mermin-Wagner theorem, which does not apply to the TC model as this is an fully connected model and the theorem does not apply here. On the other hand, since there is no inherent length scale in the TC model the whole meaning of long range order is lacking. It is interesting to note that upon eliminating the boson degrees-of-freedom of the TC model one derives the infinite range $X\!X$ model which is a special example of the Lipkin-Meshkov-Glick model,
\begin{equation}
\hat H_\mathrm{LMG}=\frac{\Omega}{2}\hat S_z+\frac{\lambda}{S}\left(\hat S_x^2+\hat S_y^2\right)=\frac{\Omega}{2}\hat S_z+\frac{\lambda}{S}\left(S(S+1)-\hat S_z^2\right).
\end{equation}
Clearly, the eigenstates are those of $\hat S_z$, {\it i.e} $|S,m_s\rangle$, and up to the critical point the ground state is $|S,-S\rangle$. Beyond the critical point, the $m_s$ quantum number increases monotonically with $g$, and in the thermodynamic limit the magnetization $\langle\hat S_z\rangle$ becomes smooth. In Tab.~\ref{tab1} we summarize some of the results of the NS PT's studied for the different models.
  
Following Refs.~\cite{rabipt1,rabipt2,plenio1,plenio4} we also analyzed the so called classical limit of the quantum Rabi and the JC models. In this limit the two models become critical, and there are many similarities with the critical behaviour of the Dicke and TC models. For example, the PT's are again of the mean-field type and the critical exponents the same. Even though the universality are the same for the Dicke and quantum Rabi as for the TC and JC models, we pointed out that criticality in the Rabi and JC models appear only in the extreme detuning. 

Allowing for photon losses is most relevant when discussing optical realizations of the the NS PT's.
PT's for these open quantum systems are defined via the properties of the steady state $\hat\rho_\mathrm{ss}$. For the Rabi and Dicke models, transitions between the different $\mathcal K$-sectors are driven by the counter rotating terms, and  similarly the photon loss processes also induce transitions between the sectors. The steady state is then the one when these transitions are balanced, which is different from the vacuum and in general a statistical mixture of the two parity states~\cite{joshi}. The criticality of these two models survives the loss of photons, and some critical exponents remain the same as studied here. However, other exponents do change once photon losses are included~\cite{critDicke}. The situation is qualitatively different for the TC and JC models since there are no counter rotating terms causing transitions between the $\mathcal K$-sectors. Photon losses only permits for transitions between the sectors by lowering the excitation number $K$ and eventually the system ends up in the vacuum $\hat\rho_\mathrm{ss}=|0,-N/2\rangle\langle0,-N/2|$ regardless of parameters and initial state. Thus, the open TC and JC models are {\bf not} critical. 

\ack
The author thanks Chaitanya Joshi, Sahel Ashhab, Emanuele Della Torre, Andr\'as Vukics for helpful discussions, and acknowledges financial support from the Knut and Alice Wallenberg foundation and the Swedish research council (VR).  

\section*{Appendix}
In Sec.~\ref{sec2} we summarized the basics of the transverse Ising model in one dimension. Those results were later put in connection with the PT's of the Dicke and Rabi models as breaking of a $\mathbb{Z}_2$ symmetry. The TC and JC models supports, however, a continuous $U(1)$ symmetry, and as an example of a model supporting a continuous symmetry we look at the $X\!X$ model in one dimension
\begin{equation}\label{XX}
\hat H_{X\!X}=-\sum_i\left[\hat\sigma_i^z+g\left(\hat\sigma_i^x\hat\sigma_{i+1}^x+\hat\sigma_i^y\hat\sigma_{i+1}^y\right)\right].
\end{equation}
This model is invariant under rotations around the $z$-spin axis. After application of the Jordan-Wigner transformation, the model is diagonalized in the momentum representation~\cite{sachdev}
\begin{equation}\label{xx2}
\hat H_{X\!X}=\sum_k\varepsilon_k\hat c_k^\dagger\hat c_k,
\end{equation}
with the dispersion 
\begin{equation}\label{xx3}
\varepsilon_k=-2g\cos(k)-1.
\end{equation}
For $g^{-1}<-2$ the dispersion is positive so the ground state is the vacuum, while for $g^{-1}>2$ it is completely negative and all momentum states are occupied. In between there is a smooth transition where the states are filled up for increasing $g$. By dividing the expression (\ref{xx3}) by $g$ we can think of the second term $1/g$ as a chemical potential that determines the fermi level and thereby the number of particles. The ground state energy is $E_0(g)=\int_{k^-}dk\,\varepsilon(k)$ where by $k^-$ we mean all momenta such that $\varepsilon(k^-)<0$. It is clear that the spectrum is gapless (Goldstone mode) and that $E_0(g)$ is a continuous function. In the fermion representation, the rotational symmetry of the model translates into conservation of particle number, similar to what we found for the TC and JC models. Importantly, given a coupling $-2<g^{-1}<2$, as $g^{-1}$ ({\it i.e.} the effective chemical potential) is varied within this interval the number of particles for the ground state changes, but since the number of particles is conserved, the corresponding eigenenergies intersect in true crossings. Thus, as for the TC and JC models there are no quantum fluctuations behind the transition. For a finite system the quasi momentum is discrete $k_n=-\pi/2+n\pi/N$ with $n=1,\,2,\,...,\,N$. It is clear that in this situation the crossings of the ground state energies line up in a series (by increasing the 'chemical potential' $g^{-1}$ more and more momentum states become populated), similar as for the TC and JC models. When taking the thermodynamic limit, the quasi momentum becomes continuous and the crossings merge into one critical point, and for $-2<g^{-1}<2$ the spectrum is gapless.

There are spectral similarities between the $X\!X$ model and the TC and JC ones with the true crossings clustering into a single critical point, and moreover the ground state energy for the $X\!X$ model has a discontinuity in $\partial_g^2E_0(g)$ at that point. Recalling the Mermin-Wagner theorem~\cite{AA}, however, we know that in one dimension a continuous PT is not allowed, and thus, the transition of the $X\!X$ model is rather a {\it Kosterlitz-Thouless} PT~\cite{RG}.

\end{document}